\documentclass[preprint,aps,prb,showpacs]{revtex4}
\usepackage{amsmath}
\usepackage{amsfonts}
\usepackage{amssymb}
\usepackage{graphicx}
\begin{document}
\title{Theory of Low-Temperature Hall Effect in Stripe--Ordered  Cuprates}
\author{Jie Lin and A. J. Millis}
\affiliation{Department of Physics, Columbia University, 
538 West 120th Street, New York, NY 10027}

\begin{abstract}
We investigate the effect of static 
anti--phase stripe order on the weak--field Hall 
effect of electrons 
on a two--dimensional square lattice with electron dispersion 
appropriate to the high T$_c$ cuprates. 
We first consider the cases where the magnitudes of the 
spin and charge stripe potentials are smaller than or 
of the same order as the bandwidth of the 
two--dimensional electrons, so that the electronic 
properties are not too strongly one--dimensional. 
In a model with only spin stripe potential, and at carrier 
concentrations appropriate to hole--doped cuprates, increasing 
the stripe scattering potential from zero leads to an increase 
in $R_H$, followed by a sign change. If the scattering 
amplitude is yet further increased, 
a second sign change occurs. The results 
are in semiquantitative agreement with data. 
In a charge--stripe--potential--only model, 
$R_H$ increases as the charge stripe scattering 
strength increases, with no sign change occurring. 
In a model with both spin and charge stripe potentials, 
$R_H$ may be enhanced or may change sign, depending on the 
strengths of the two scattering potentials. We also consider 
the case in which the magnitudes of the stripe potentials are much 
larger than the bandwidth, where analytical results 
can be obtained. In this limit, the system is 
quasi--one--dimensional, while $R_H$ remains finite and its sign 
is determined by the carrier density and the 
electron band parameters. 
\end{abstract}
\maketitle

\section{Introduction}

Stripe order, static or fluctuating, is argued to be 
an important ingredient in understanding the physics of the high 
temperature superconductors.\cite{Orenstein00,Kivelson03} 
In the YBa$_2$Cu$_3$O$_{6+x}$ family, stripe order was 
recently used to explain \cite{Millis07} the small electron 
pockets observed in the quantum oscillation measurements. 
\cite{Doiron-Leyraud07,Bangura08} 
In the family of materials derived from La$_2$CuO$_4$, stripe 
order is believed to be prevalent, being related to the 
``1/8 anomaly'' observed in most members of this material family.%
\cite{Moodenbaugh88} 
In the La$_{1.6-x}$Nd$_{0.4}$Sr$_x$CuO$_4$ (Nd--LSCO) series,
static stripe order has been shown by neutron diffraction 
measurements to exist over a significant part of the 
temperature--doping phase diagram,\cite{Ichikawa00} 
up to Sr doping $x\approx0.25$. 

The Hall resistance of 
La$_{1.6-x}$Nd$_{0.4}$Sr$_{x}$CuO$_{4}$ systems 
has been studied experimentally.\cite{Daou08,Nakamura92} 
It was found that at the nominal hole doping $x=0.24$, 
the low temperature Hall coefficient $R_H$ takes the value appropriate to 
a two--dimensional metal with carrier (hole) density $1+x$. 
However, for the lower dopings $x=0.20$ and $x=0.12$, 
the measured $R_H$ deviates significantly from 
what is expected for a conventional metal with 
carrier density $1+x$. At $x=0.20$, $R_H$, while positive, is 
much larger than the value expected from the conventional model. 
For $x=0.12$, the sign of $R_H$ is opposite, 
showing an electron--like behavior.  
A similar issue arises in the electron--doped cuprates 
Pr$_{2-x}$Ce$_x$CuO$_4$ (PCCO), \cite{Dagan04} 
where the Hall number is positive for doping $x> 0.15$ and 
becomes negative for smaller dopings. 
In the electron--doped material, the change of sign was 
explained by a commensurate $(\pi,\pi)$ 
spin density wave order.\cite{Lin05} 
However, in the hole--doped materials, 
$(\pi,\pi)$--ordering would not produce a 
sign change. 
In this paper, we investigate whether stripe order can 
account for the magnitude and 
the unconventional doping dependence of the 
Hall resistivity observed in the La/Nd-Sr-Cu-O compounds. 

The rest of this paper is organized as follows. Sec. \ref{model} 
defines a phenomenological model for band electrons in the 
presence of stripe order, and summarizes the formulae used to 
calculate the conductivities. 
Sec. \ref{FS} illustrates the evolution of the Fermi surface 
in the stripe ordered state. 
Sec. \ref{comparison} 
discusses the effects of the charge stripe potential 
and the spin stripe potential on transport properties. 
Sec. \ref{results} presents the doping
dependence of $R_H$ in the spin stripe ordered state. 
Sec. \ref{strong coupling} discusses 
the Hall effect in the strong stripe potential limit. 
Sec. \ref{discussion} is a conclusion in which the results 
are summarized and discussed and implications are outlined. 

\section{Model and Formalism}
\label{model}

We assume electrons moving on a two--dimensional 
square lattice of unit lattice constant, 
with a band dispersion given by 
\begin{equation}
\begin{split}
\varepsilon_p=&-2t(\cos p_x+\cos p_y)+4t^{\prime} \cos p_x\cos p_y \\
&-2t^{\prime\prime} (\cos 2p_x+\cos2p_y).
\label{bare dispersion}
\end{split}
\end{equation}
In our numerical calculations, 
we use the canonical values\cite{Andersen95} $t=0.38$eV, 
$t^{\prime}=0.32t$, and $t^{\prime\prime}=0.5t^{\prime}$. 
In addition, we assume that the electrons feel the effect of static 
``stripe'' (spin and charge density wave) order. Because 
we are interested only in low--temperature transport, we 
neglect fluctuations and treat the order in the mean--field 
approximation. 

We take the spin modulation to be longitudinal 
and to be described by 
the wave vector ${\bf Q}_s$, 
so that it gives rise to the scattering 
potential 
\begin{equation*}
 \Delta_s({\bf R})=2V\cos {\bf Q}_s\cdot{\bf R}. 
\end{equation*}
The spatial periodicity of 
this potential can be obtained from the incommensurate peaks 
in neutron diffraction measurements. 
Tranquada {\it et.al.}\cite{Tranquada97} 
showed that in the Nd--LSCO series for $x\lesssim 1/8$, 
${\bf Q}_s=\pi(1-2x,1)$, 
while for $x>1/8$, the spin incommensurability is 
approximately doping independent, with wave vector ${\bf Q}_s^{\ast} 
\approx \pi(3/4,1)$. We will be mainly interested in doping $x>1/8$, so 
we fix ${\bf Q}_s={\bf Q}_s^{\ast}$. 

Charge modulations 
are also observed in the Nd--LSCO materials.\cite{Tranquada95} 
These occurs at the wave vector ${\bf Q}_c=2{\bf Q}_s$ 
expected from general 
Landau theory arguments, which allow a term $S_Q^2\rho_{-2Q}$ in the 
free energy, where $S_Q$ and $\rho_{-2Q}$ are the 
spin stripe and charge stripe order parameters, 
respectively.\cite{Zachar98} 
We model the effect of charge stripes by the potential, 
\begin{equation*}
 \Delta_{Q}({\bf R})=2V_c\cos {\bf Q}_c\cdot{\bf R}, 
\end{equation*}
and we set ${\bf Q}_c=2{\bf Q}_s^{\ast}=\pi(1/2,0)$. 
We have approximated the stripe potentials as simple cosines; 
deviations from this form were investigated and found 
not to be important.\cite{Millis07} 

These considerations 
lead to the following Hamiltonian: 
\begin{equation}
H=\left(\begin{array}{cccccccc}
\varepsilon_p & V_c & 0 & V_c & 0 & V & V & 0\\
V_c & \varepsilon_{p+(\frac{1}{2},0)\pi} & V_c & 0   & 0 & 0 & V & V\\
0 & V_c & \varepsilon_{p+(1,0)\pi}     & V_c & V   & 0 & 0 & V\\
V_c & 0 & V_c & \varepsilon_{p+(\frac{3}{2},0)\pi} & V & V & 0 & 0 \\
0 & 0 & V & V & \varepsilon_{p+(\frac{1}{4},1)\pi} & V_c & 0 & V_c \\
V & 0 & 0 & V & V_c & \varepsilon_{p+(\frac{3}{4},1)\pi} & V_c & 0 \\
V & V & 0 & 0 & 0 & V_c  & \varepsilon_{p+(\frac{5}{4},1)\pi}& V_c \\
0 & V & V & 0 & V_c & 0 & V_c & \varepsilon_{p+(\frac{7}{4},1)\pi}
\end{array}\right).
\label{hamiltonian}
\end{equation}

We assume that the low temperature DC 
transport can be described by the Boltzmann equation. 
We further assume, as is appropriate for 
low temperatures, that the relaxation is mainly due to randomly 
distributed impurities with a low density, leading to 
a constant scattering rate, $1/2\tau$. 
The expressions for the longitudinal and Hall conductivities 
then follow from solving the Boltzmann equation with 
the relaxation time approximation 
(for a detailed derivation, see Ref. [\onlinecite{Lin05}]). 
Assuming the $T\to 0$ limit can be taken, these expressions are 
one--dimensional integrals along the Fermi surface, 
\begin{align}
\sigma_{xx}&=\frac{\sigma_Q}{4\pi^2}\tau
\oint ds \frac{v_x(s)^2}{v_F(s)},\label{sxx}\\
\sigma_{yy}&=\frac{\sigma_Q}{4\pi^2}
\tau\oint ds \frac{v_y(s)^2}{v_F(s)},\label{syy}
\end{align} 
\begin{equation}
\sigma_{xy}=\sigma_Q\frac{B}{\Phi_0}\frac{1}{4\pi}
\tau^2\oint {\bf v}\times d{\bf v}\cdot\hat{\bf z}, 
\label{sxy}
\end{equation}
where $s$ is the arc length coordinate along the 2D Fermi surface,  
$\hat{\bf z}$ is the unit vector along the $c$--axis, 
and ${\bf v}$ is the Fermi velocity. In these equations, 
$\sigma_Q=e^2/\hbar$ 
is the conductance quantum, and 
$\Phi_0=hc/2e$ is the superconducting flux quantum. 
The Hall coefficient $R_H=\sigma_{xy}/(B\sigma_{xx}\sigma_{yy})
=1/nec$ with $n$ an effective electron density per unit cell 
per plane. 

We evaluate these equations by first identifying 
the bands which produce Fermi surface segments, 
then using a numerical search procedure to locate 
the Fermi surface. Typically, $\sim 10^4$ Fermi 
surface points are used. We then compute the 
velocities at each point and evaluate the integrals 
by the trapezoidal rule.

\section{Fermi Surface Evolution in the Spin and Charge 
Stripe Ordered States}
\label{FS}

\begin{figure}[htbp]
 \centering
\includegraphics[width=3in]{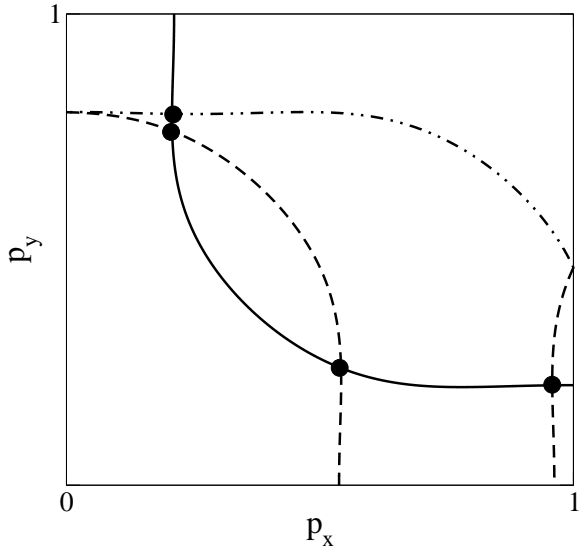}
\caption{\small{
The normal state Fermi surface (solid line) for 
doping $x=1/8$ and its translations 
by ${\bf Q}_s=\pi(3/4,1)$ (dashed line) 
and by $(2\pi,2\pi)-{\bf Q}_s=\pi(5/4,1)$ (dashed--dotted line) 
in the first quadrant of the first Brillouin zone. 
The 4 hot spots are shown here as solid points.
In this and the following Fermi surface 
plots, the unit of momentum $p$ is $\pi/a$, 
with $a=1$ the lattice constant 
of the square lattice.}}
\label{normalFS}
\end{figure}

As shown in Ref. [\onlinecite{Millis07}], in 
the mean--field stripe ordered state the electron 
Fermi surface is reconstructed from the one obtained 
in the band theory calculation 
in a complicated way. 
The normal state Fermi surface for doping $x=1/8$ 
is shown as the solid line in Fig. [\ref{normalFS}], along with 
its translations by ${\bf Q}_s=\pi(\frac{3}{4},1)$ (dashed line)
and by $(2\pi,2\pi)-{\bf Q}_s=\pi(\frac{5}{4},1)$ 
(dashed--dotted line). 
For small $V$ and/or $V_c$, reconstruction happens 
in the vicinity of the hot spots (shown as solid points 
in Fig. [\ref{normalFS}]), where the Fermi surface crosses 
itself upon translation by the stripe wave vectors.  
In Fig. [\ref{normalFS}], we only show two values of the 
stripe wave vectors for simplicity. The complete Fermi surface 
crossing can be found 
in Ref. [\onlinecite{Millis07}]. 

\begin{figure}[htbp]
 \centering
\includegraphics[width=6in]{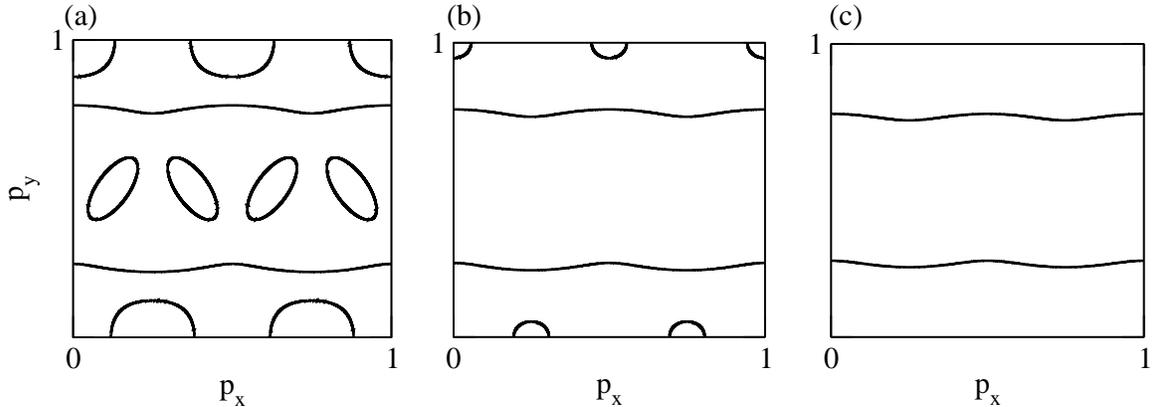}
\caption{\small{Fermi surface evolution 
in the spin stripe ordered state, with 
the charge stripe potential $V_c=0$ and doping $x=1/8$. 
(a): $V=0.2$eV. (b): $V=0.25$eV. (c): $V=0.3$eV.}}
\label{changeVfs}
\end{figure}

The Fermi surface evolution in the absence of the charge 
stripe potential is illustrated in Fig. [\ref{changeVfs}], 
where the Fermi surfaces are plotted from left to right 
for increasing values of $V$. We see from 
Fig. [\ref{changeVfs}(a)] that 
at relatively small $V$, there are 
well--defined hole pockets centered at 
$(\pm \pi/8,\pi/2)$, electron pocket centered at $(0,\pi)$
and open Fermi surface. 
When $V$ is increased further, the hole pockets are eliminated, 
Fig. [\ref{changeVfs}(b)], and 
at a still larger $V$, the electron pocket is eliminated, leaving 
the open Fermi surface alone, Fig. [\ref{changeVfs}(c)].

\begin{figure}[htbp]
 \centering
\includegraphics[width=5in]{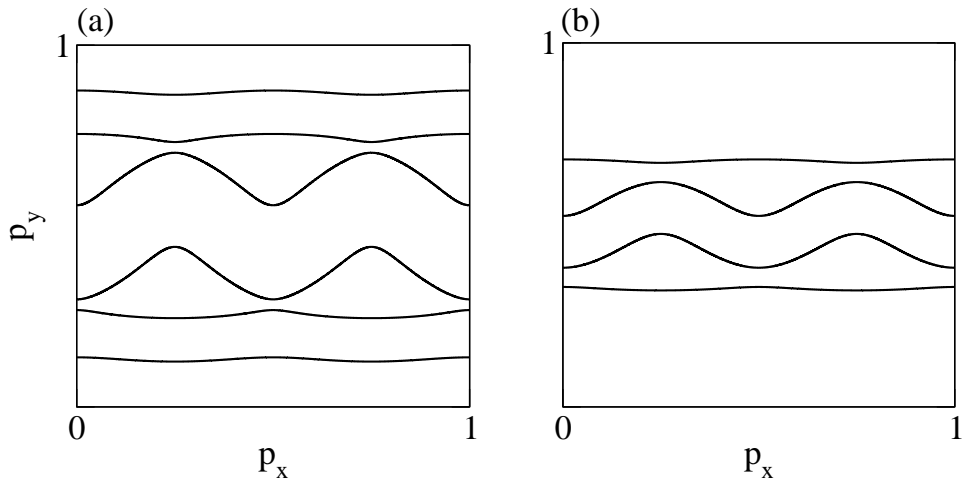}
\caption{\small{Fermi surface in the charge stripe ordered state, 
with the spin stripe potential $V=0$ and a fixed doping $x=1/8$. 
(a): $V_c=0.15$eV. (b): $V_c=0.35$eV.}}
\label{changeVCfs}
\end{figure}

The Fermi surface evolution due to a charge stripe potential 
in the absence of the spin stripe potential 
is plotted in Fig. [\ref{changeVCfs}]. We see that for the two 
$V_c$ values shown here, the Fermi surface is open, 
with no pockets. 

\begin{figure}[htbp]
 \centering
\includegraphics[width=3in]{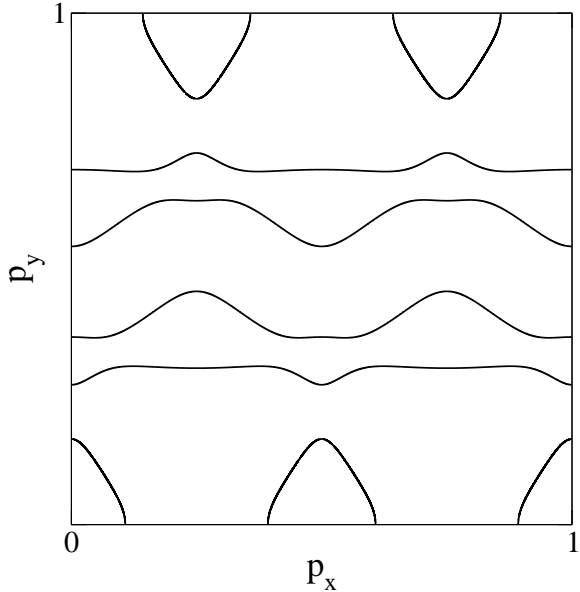}
\caption{\small{The Fermi surface in the presence 
of both spin and charge stripe orders 
at doping $x=1/8$, $V=0.2$eV and $V_c=0.3$eV.}}
\label{vsvcfs}
\end{figure}

When both types of stripe potentials are 
present, the Fermi surface reconstruction is more complicated. 
One representative Fermi surface is shown in Fig. [\ref{vsvcfs}]. 
In the case plotted, three bands cross the Fermi level. 
Two of them give open Fermi surfaces, 
while the third one gives an electron pocket 
centered at $(0,0)$. 

The consecutive changes of the Fermi 
surface topology upon changing the stripe potentials 
influence the Hall conductivity $\sigma_{xy}$, 
and the longitudinal conductivities $\sigma_{xx}$ and $\sigma_{yy}$, 
and will be studied in the next section.

\section{Hall Effect: spin stripe potential vs. 
charge stripe potential}
\label{comparison}

\subsection{Overview}

In this section, we will consider the Hall effect 
in the spin and/or charge 
stripe potentials. Separate subsections treat different cases. 
We shall first consider the case where both potentials 
are weak, that is, the system is close to the quantum critical point 
from the normal state to the stripe ordered state. Then, we consider 
the cases of spin stripe potential only 
and charge stripe potential only, and finally the effects 
of combined spin- and charge- stripe scattering.
In this section, we shall fix the doping to be $x=0.125$, where 
the stripe order is most stable, and study the Hall effect, 
changing the strength of the stripe potentials $V$ and $V_c$. 
In section \ref{results}, we treat the doping dependence. 

\subsection{Critical Behavior close to the Stripe Order 
Quantum Phase Transition}

At small $V$ and/or $V_c$, the Fermi surface 
reconstructs in the vicinity 
of the hot spots; it is essentially unchanged far from those points. 
Thus, although there are several Fermi surface crossings 
due to the $8\times 8$ matrix structure of $H$, 
the total changes in $\sigma_{xy}$, 
$\sigma_{xx}$ and $\sigma_{yy}$ are additive. 
For each Fermi surface crossing, our 
previous analysis\cite{Lin05} applies. 
We find $\delta\sigma_{xy}$ and 
$\delta\sigma_{xx}+\delta\sigma_{yy}$ are both linear in $V$ and $V_c$, 
such that as $V\to 0$ and $V_c\to 0$, 
\begin{equation}
 \frac{\delta R_H}{R_H^0}=\frac{\delta\sigma_{xy}}{\sigma_{xy}^0}
-\frac{\delta\sigma_{xx}+\delta\sigma_{yy}}{\sigma_{xx}^0}
=aV+bV_c,
\label{asymptotic}
\end{equation}
where the superscript 0 denotes the corresponding value in the 
normal state, and we have used the fact that the normal state 
has 4--fold symmetry, so that $\sigma_{xx}^0=\sigma_{yy}^0$. 
The prefactors $a$ and $b$ can be determined; 
$a\approx6$eV$^{-1}$, and $b\approx10$eV$^{-1}$. 
In the case of PCCO,\cite{Lin05} this 
asymptotic formula holds only within 1\% of the critical value. The more 
complicated Fermi surfaces found here will restrict the domain 
of validity even further. For the very small values of 
$V$ and $V_c$ for which Eq. [\ref{asymptotic}] applies, 
there are complications due to strong field 
crossover\cite{Fenton05} or magnetic breakdown.\cite{Blount62}

\subsection{Spin Stripe Potential Only}

\begin{figure}[htbp]
 \centering
\includegraphics[width=5in]{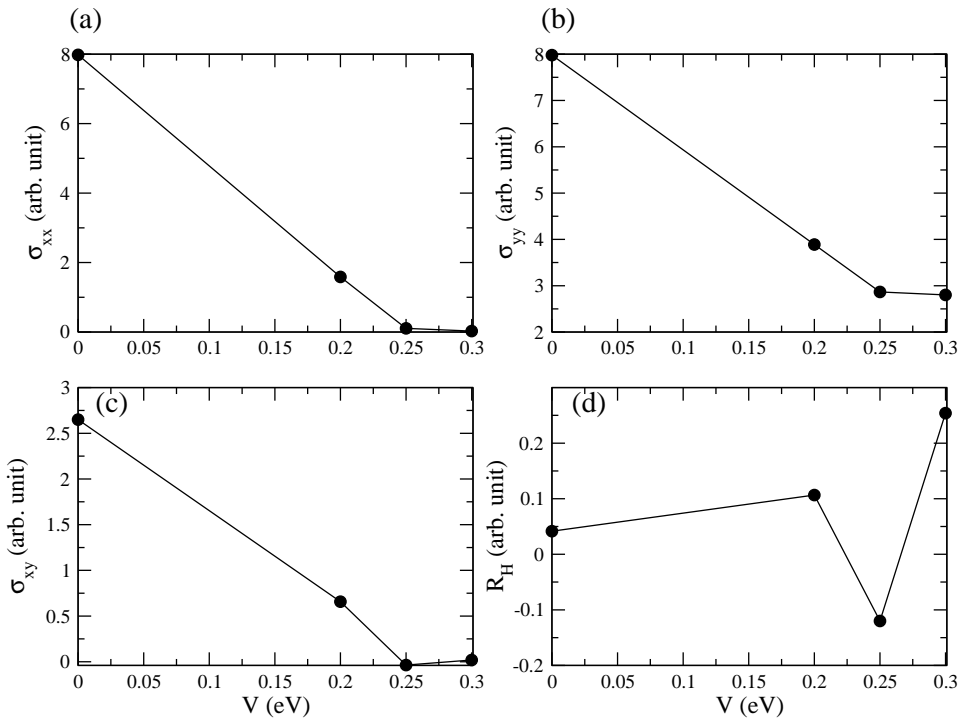}
\caption{\small{Transport coefficients in the spin stripe ordered state at 
doping $x=1/8$ and $V_c=0$. 
(a): $\sigma_{xx}$ as a function of $V$. 
(b): $\sigma_{yy}$ as a function 
of $V$. (c): $\sigma_{xy}$ as a function of $V$. 
(d): $R_H$ as a function of $V$. 
Here and in the following plots of the transport properties, 
the solid lines are guides to the eye; exceptions 
will be stated explicitly. }}
\label{changeV}
\end{figure}

In this subsection, we consider the case where the electrons are 
scattered only from the spin stripe potential, namely, 
$V_c=0$. For each value of $V$, we calculate  
the loci of the Fermi surface and the Fermi velocities, and 
use this information in 
Eqs. [\ref{sxx}--\ref{sxy}]. Fig. [\ref{changeV}] 
shows the results of such a calculation. We note that 
the onset of the stripe potential suppresses both 
the longitudinal and the Hall conductivities, 
Fig. [\ref{changeV}(a)--(c)], 
as in the case of the commensurate spin density wave 
order in PCCO.\cite{Lin05} Fig. [\ref{changeVfs}] shows the 
corresponding Fermi surfaces for the three nonzero $V$ values. 
We can see that the behavior of $\sigma_{xy}$ can be 
qualitatively understood in terms of the evolution of 
the Fermi surface topology.\cite{Ong91}
For $V=0.2$eV, $R_H$ is positive and significantly 
larger than the value in 
the normal state. The sign is due to the dominant hole pockets 
(Fig. [\ref{changeVfs}(a)]), and 
the enhancement is due to the strong decrease of 
the longitudinal conductivities. 
When $V$ grows large enough (0.25eV here), $R_H$ becomes negative. 
This reversal of sign 
comes from the elimination of the hole pockets 
(Fig. [\ref{changeVfs}(b)]). 
The smallness of the open Fermi surface 
contribution to $\sigma_{xy}$ means that the electron 
pocket determines the sign of $R_H$. At still larger $V$ (0.3eV), 
the electron pocket is eliminated 
(Fig. [\ref{changeVfs}(c)]). The open Fermi surface 
gives rise to a small positive contribution to $\sigma_{xy}$ 
(hole--like), so $R_H$ becomes positive again. 
However, the sign of $R_H$ due to the open Fermi surface 
changes as $V$ is increase further. As will be shown 
in Sec. \ref{strong coupling}, the crossover to the 
strong coupling limit occurs at $V\sim 1$eV. 

We notice that in the cases with $V=0.25$eV and 0.3eV, $|R_H|$ is 
quite large compared to the band value $R_H(V=0)$, although 
$\sigma_{xy}$ is much smaller than $\sigma_{xy}(V=0)$. 
This is due to the large anisotropy, 
as measured by $\sigma_{yy}/\sigma_{xx}$, for these two $V$ values. 
This anisotropy is due to the open Fermi surfaces and 
grows rapidly with increasing $V$. This anisotropy compensates for the 
smallness of $\sigma_{xy}$, giving a large $R_H$.

Fig. [\ref{changeV}(d)] shows $R_H$ for the 3 values of $V$. 
We see that as $V$ grows, 
$R_H$ first increases such that at $V=0.2$eV, 
$R_H/R_H^0\approx 2.5$, and then decreases to a negative value. This 
trend is qualitatively consistent with the experimental data. In the 
next section, we shall study the doping dependence of $R_H$, 
assuming a model in which the spin stripe potential 
opens at $x=0.24$ and grows as doping is reduced. 
We shall see that such a model can semi--quantitatively 
account for the experimental data.

\subsection{Charge Stripe Potential Only}

\begin{figure}[htbp]
 \centering
\includegraphics[width=5in]{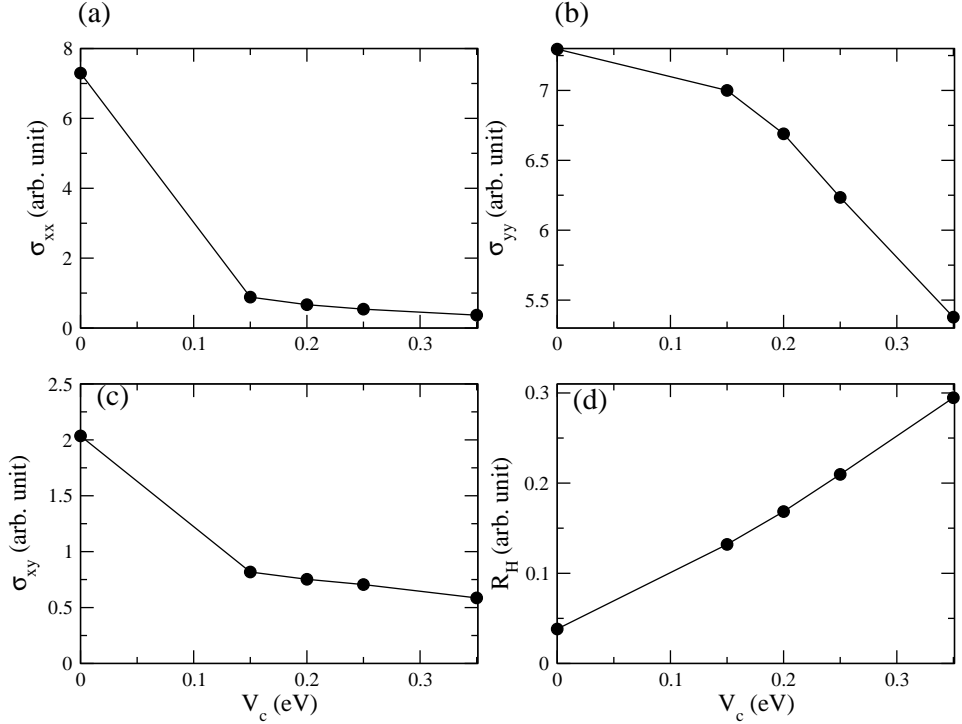}
\caption{\small{Transport coefficients 
in the charge stripe ordered state 
calculated for doping $x=1/8$ and $V=0$. 
(a): $\sigma_{xx}$ as a function of $V_c$. 
(b): $\sigma_{yy}$ as a function 
of $V_c$. (c): $\sigma_{xy}$ as a function of $V_c$. 
(d): $R_H$ as a function 
of $V_c$.}}
\label{changeVC}
\end{figure}

We now consider the effects of $V_c$ on the transport properties 
with the spin potential set to 0, Fig. [\ref{changeVC}].  
Fig. [\ref{changeVCfs}] shows 2 representative 
Fermi surfaces for $V_c=0.15$eV and $V_c=0.35$eV.  
For $V_c$ in this range, all the pieces of the Fermi 
surface are open. However, some pieces of the 
open Fermi surface have relatively large curvature, 
because they can be viewed as the merging of 
the hole Fermi pockets. 
As shown in Fig. [\ref{changeVC}(c)], $\sigma_{xy}$ is thus  
always relatively large even for $V_c=0.35$eV.  From 
Fig. [\ref{changeVC}(d)], we see that the onset of the 
charge stripe order gives a rapid increase of $R_H$. In fact, 
$R_H(V=0,V_c=0.2{\rm eV})=1.5 R_H(V=0.2{\rm eV},V_c=0)$. 
We also notice that the anisotropy 
is less than that in the spin stripe case; 
in the spin stripe case, 
$\sigma_{yy}/\sigma_{xx}\approx 100$ at $V=0.3$eV, 
while in the charge stripe case, 
$\sigma_{yy}/\sigma_{xx}\approx 15$ at $V_c=0.35$eV.  

We see that although the increase of the charge stripe potential 
substantially enhances $R_H$, $R_H$ remains positive for 
all the four $V_c$ values considered here. 
Further calculation (not shown here) suggests that $R_H$ 
changes sign around $V_c=0.8$eV, and approaches a 
negative value in the limit $V_c\gg t$. 
This suggests that a model with only charge stripe order 
is inconsistent with experimental data.

\subsection{Coexistence of the Spin Stripe Potential 
and the Charge Stripe Potential}

\begin{figure}[htbp]
 \centering
\includegraphics[width=5in]{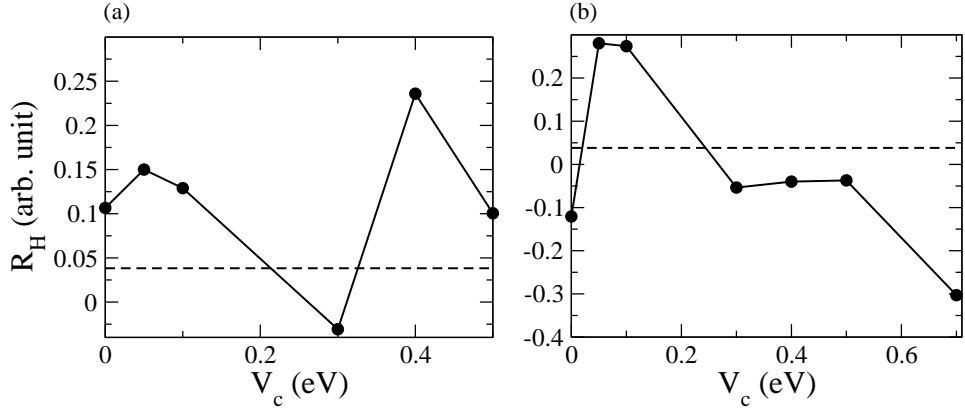}
\caption{\small{$R_H$ for nonzero $V$ and nonzero $V_c$ 
and doping $x=1/8$. (a): $V=0.2$eV; (b): $V=0.25$eV. 
Dashed line indicates the position of $R_H^0$.}}
\label{vsvc}
\end{figure}

Now we study the case in which the 
spin stripe and the charge stripe coexist, 
$V\neq 0$ and $V_c\neq 0$. The 
interplay between these two stripe potentials leads 
to very complicated behavior of $R_H$. 
Fig. [\ref{vsvc}] shows two representative sets of results.
In both cases, we fix $V$ and increase $V_c$ from 0 to a 
large value. 
Fig. [\ref{vsvc}(a)] shows $R_H (V_c)$ for $V=0.2$eV. 
In this case, $R_H(V_c=0)>0$, and at a small $V_c=0.05$eV, 
$R_H$ is strongly enhanced by a factor of $1.5$.  
For $V_c=0.3$eV, $R_H$ becomes negative, while at larger 
$V_c$, $R_H$ is positive again. Fig. [\ref{vsvc}(b)] shows 
$R_H$ as a function of $V_c$ for $V=0.25$eV. In this case, 
$R_H(V_c=0)<0$. 
At a small value of $V_c=0.05$eV, 
the sign is reversed to be hole--like, 
while at larger $V_c$, $R_H$ becomes negative again.  

The behavior of $R_H$ can be understood 
in terms of the evolution of 
the Fermi surface topology when changing $V_c$ and $V$. 
In the case of $V=0.2$eV, 
for both $V_c=0.05$eV and $V_c=0.1$eV, the 
calculated Fermi surface (not shown) closely resembles that in 
Fig. [\ref{changeVfs}(a)], explaining the positive sign of $R_H$. 
For the case $V_c=0.3$eV, 
the Fermi surface is given in Fig. [\ref{vsvcfs}] and has 
a qualitatively different topology, such that 
the hole Fermi surface becomes open, 
and the electron pocket (which now dominates $\sigma_{xy}$) 
changes from being centered at $(0,\pi)$ to being centered 
at $(0,0)$. At larger values of $V_c$, 
the electron pocket is eliminated, leaving open Fermi surface only, 
qualitatively resembling Fig. [\ref{changeVCfs}(b)].  
The sign of $\sigma_{xy}$ is calculated to be positive. 

In the case of stronger $V=0.25$eV, 
the $V_c=0$ Fermi surface is shown in Fig. [\ref{changeVfs}(b)]. 
At $V_c=0.05$eV, the 
hole pockets reappear, with a very small radius, 
leading to a Fermi surface very similar to that in 
Fig. [\ref{changeVfs}(a)]. However, the small 
hole pockets dominate the sign of $R_H$. 
The hole Fermi pockets grow with increasing $V_c$, 
and eventually merge into open Fermi surface at large $V_c$. 
For $V_c=0.3$eV, the Fermi surface topology changes 
qualitatively as in the case of $V=0.2$eV and $V_c=0.3$eV, 
Fig. [\ref{vsvcfs}]. This Fermi surface reconstruction 
can qualitatively explain the negative sign of $R_H$ at 
$V_c=0.3$ eV. At $V_c=0.5$ eV and 0.7 eV, 
the electron pocket is eliminated, leaving 
only open Fermi surface, as qualitatively represented 
in Fig. [\ref{changeVCfs}(b)]. 
However, the sign of $R_H$ remains negative. 

The discussion in this section shows that the interplay between 
the spin and the charge stripe potentials leads to 
two possibilities to account for the experimental observation of 
the sign change of $R_H$. In the simplest case, the spin stripe 
order is dominant, and the charge stripe order potential is small; 
$V_c$ should be less than 0.05eV when $V=0.25$eV and $x=1/8$. 
Then we assume that $V_c$ can be neglected. In the other scenario, 
both $V$ and $V_c$ are large, as shown in Fig. [\ref{vsvc}]. 
In the next section, we pursue the first scenario in more detail.

\section{Hall effect: doping dependence}
\label{results}

We now study the doping dependence of the transport coefficients. 
Doping has two effects, changing the 
carrier density and changing the strength of the stripe potential. 
From the discussion of the last section, we assume a model for 
the electrons in the stripe ordered state in which 
the spin stripe scattering is dominant, and the charge stripe 
scattering is neglected. 

\begin{figure}[htbp]
 \centering
\includegraphics[width=3in]{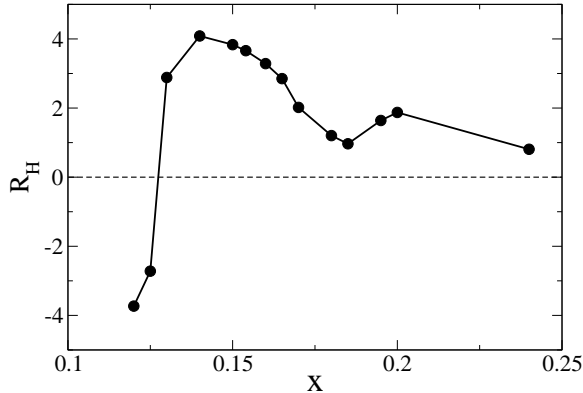}
\caption{\small{$R_H$, expressed as the inverse of 
the effective carrier density per plane per cell, 
as a function of doping. The spin stripe 
potential $V$ takes a mean field form $V{\rm [eV]}=0.35
\sqrt{1-x/0.24}$, and the charge stripe potential 
is neglected, $V_c=0$. $R_H(x=0.24)\approx 0.8\approx 1/(1+x)$. 
Dashed line indicates $R_H=0$.}}
\label{rhplot}
\end{figure}

We assume a mean field dependence of the stripe order parameter 
on doping, for $x<0.24$
\begin{equation}
V=V_0\sqrt{1-x/0.24}, 
\label{veq}
\end{equation}
and $V=0$ for $x>0.24$, where $V_0$ controls the rate at which the 
stripe order is setting in. Experimental results show 
that the $x=0.12$ sample has a negative $R_H$. 
Fig. [\ref{changeV}(d)] suggests that $V_0$ should be relatively large, 
such that $0.2<V(x=0.12)<0.3$eV. Thus, we 
choose $V_0=0.35$eV. Then for each doping $x$, 
the conductivities and the Hall coefficient can be calculated from 
Eqs. [\ref{sxx}--\ref{sxy}]. 
The results are shown in Fig. [\ref{rhplot}]. We observe that 
$R_H$ starts at $x=0.24$ ($V=0$) at the band value $1/(1+x)$, 
increases as doping is decreased, and jumps 
to a negative number around $x=0.13$. 

In terms of the Fermi surface evolution, for doping $x=0.12$, 
the Fermi surface can be represented 
by Fig. [\ref{changeVfs}(b)], for doping 
$x=0.125$, the Fermi surface resembles that 
in Fig. [\ref{changeVfs}(a)] with tiny 
hole pockets, and for doping in the range 
$0.13\leq x < 0.16$, the Fermi surfaces can be represented 
by Fig. [\ref{changeVfs}(a)]. 

\begin{figure}[htbp]
 \centering
\includegraphics[width=5in]{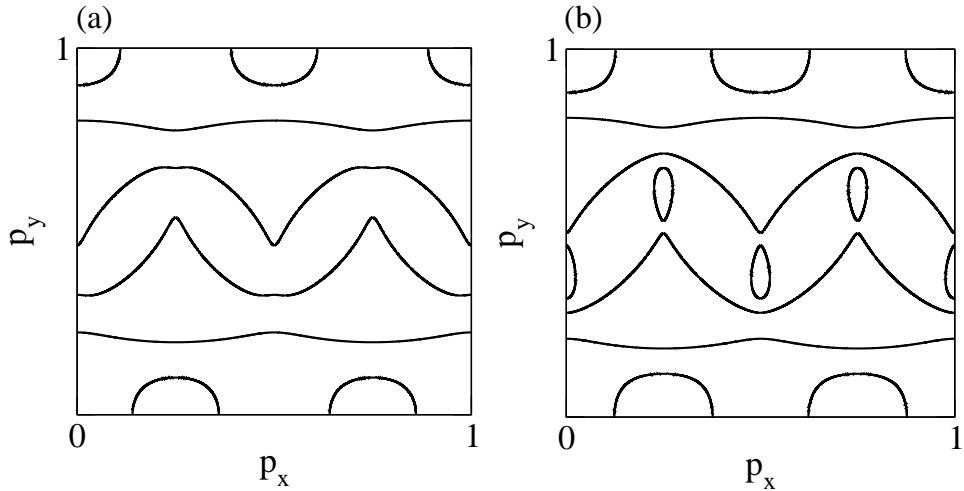}
\caption{\small{Fermi surfaces for relatively small $V$. 
(a): $V=0.175$eV, $V_c=0$, and doping $x=0.18$. 
(b): $V=0.143$eV, $V_c=0$, and doping $x=0.20$.}}
\label{interFS}
\end{figure}

We see that $R_H$ has a local minimum around $x=0.18$. 
Starting from $x=0.15$, and increasing doping (decreasing 
$V$), $R_H$ first decreases rapidly, and after $x\approx 0.18$, it 
increases, and then decreases again to the band value $R_H^0$. 
This can be qualitatively understood in terms of the 
Fermi surface evolution. At $x=0.15$, the Fermi surface 
resembles that in Fig. [\ref{changeVfs}(a)]. Increasing 
doping, the size of the hole pockets centered at $(\pm\pi/8,\pi/2)$
increases. These hole pockets eventually merge into open Fermi surface, 
see Fig. [\ref{interFS}(a)] for $x=0.18$. 
Increasing $x$ further, new pieces of Fermi surface appear, 
as shown in Fig. [\ref{interFS}(b)], which are hole--like, and 
contribute to the increase of $R_H$ around $x=0.2$. 
Because the structure of $R_H$ in the doping range $0.15<x<0.2$ 
arises from the small pockets shown in Fig.[\ref{interFS}(b)], 
we believe it will be very sensitive to details and extrinsic 
effects including scattering and magnetic breakdown. 
Increasing doping toward $x=0.24$ where $V=0$, $R_H$ is 
then described by the critical behavior, $\delta R_H/R_H^0=aV$, 
with positive $a$.

\section{Large Stripe Potential Limit}
\label{strong coupling}

\begin{figure}[htbp]
 \centering
\includegraphics[width=5in]{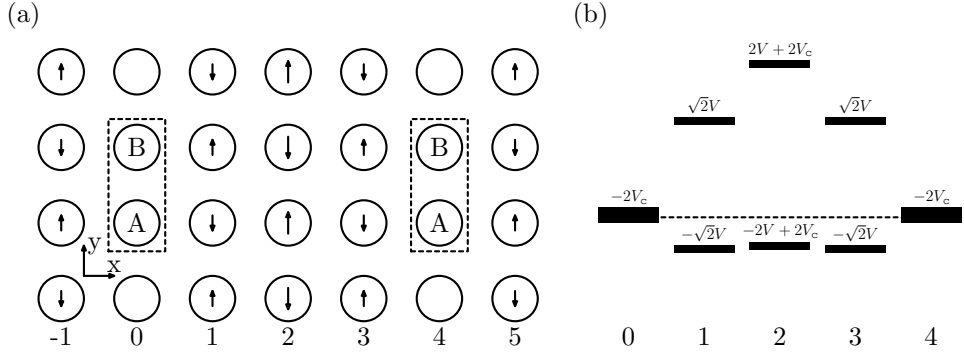}
\caption{\small{(a): the spin stripe potential for 
spin--up electrons. For spin--down electrons, 
the arrows are opposite. 
The length of arrows is proportional to the magnitude of 
the potential. The charge stripe potential is not shown here. 
(b): the energy bands for each column represented by the filled 
boxes, with height proportional to the bandwidth. In this plot, 
$V=3V_c$ and hole doping $x=0.125$. 
The Fermi level is shown as dashed line.}}
\label{RSVSVC}
\end{figure}

In this section, we consider the case where 
the stripe potential $V\gg t$ and 
the hole doping is in the range $0.125<x<0.25$. 
We assume that, on average, no two electrons 
occupy the same lattice site. We will see that 
this constraint requires that $2V_c<V$. 
Fig. [\ref{RSVSVC}(a)] shows the spin stripe 
potential for spin--up electrons. (The spin 
potential for spin--down electrons is 
opposite.) 

For large spin potentials, the 
doped holes reside on the columns 
where $\Delta_s=0$ (the circles without arrows 
in Fig. [\ref{RSVSVC}(a)]), referred to as charge stripes. 
The doped holes mainly move along these charge stripes, 
with a small probability to hop from one stripe 
to another. A general Hamiltonian for charges 
moving in weakly coupled stripes of spacing 
4 lattice constants is 
\begin{equation}
 H=\varepsilon^0(p_y)+\sum_nf_n(p_y)\cos 4np_x,
\label{general H}
\end{equation}
with small $f_n$. From Eq. [\ref{sxy}], 
the Hall conductivity is 
\begin{equation}
 \sigma_{xy}=-\mathcal{A}\int_{-\pi}^{\pi} 
dp_xv^y(p_x)\frac{dv^x(p_x)}{dp_x},
\end{equation}
where $\mathcal{A}=\sigma_Q\frac{B}{\Phi_0}\frac{1}{2\pi}
\tau^2$. At $f_n=0$, the Fermi surface is $p_y=p_0$, and 
the dispersion may be approximated by 
\begin{equation}
 \varepsilon^0(p_0+\delta p_y)=v_0\delta p_y+\frac{1}{2}
m_0\delta p_y^2.
\end{equation}
To leading order in $f_n$, we find 
\begin{equation}
 \sigma_{xy}=-16\pi\mathcal{A}\sum_nn^2
\bigl(\frac{m_0}{v_0}-\frac{1}{f_n}\frac{df_n}{dp_y}\bigr)
f_n^2,
\label{sigxy large stripe}
\end{equation}
with all quantities evaluated at $p_0$. In most cases we find 
that the sign of the Hall effect is determined by the 
curvature of the 1D band ($m_0/v_0$); but for some particular 
parameter values, structure in the interchain hopping can 
produce a sign change, for example, when one of the 
$f_n$ goes through zero. 

The longitudinal conductivities $\sigma_{xx,yy}$ can be 
calculated in a similar manner in terms of 
$f_n$. To leading order in $f_n$, 
Eq. [\ref{sxx}] can be 
approximated as 
\begin{equation}
 \sigma_{xx}=\mathcal{B}\int_{-\pi}^{\pi} dp_x\sqrt{1+
\bigl(\frac{dp_y}{dp_x}\bigr)^2}(v^x)^2/
\sqrt{(v^x)^2+(v^y)^2}
\approx \mathcal{B}\frac{16\pi}{|v_0|} 
\sum_nn^2f_n^2
\label{sigxx large stripe}
\end{equation}
where $\mathcal{B}=\sigma_{Q}\tau/(4\pi^2)$ and 
$f_n$ is evaluated at $p_0$. 
Eq. [\ref{syy}] can be approximated as 
\begin{equation}
 \sigma_{yy}\approx \mathcal{B}\int_{-\pi}^{\pi}
dp_x |v_0|=\mathcal{B}2\pi|v_0|.
\label{sigyy large stripe}
\end{equation}
In the rest of this section, 
we present an evaluation of $f_n$ in the strong 
coupling limit using perturbation theory. 

The zeroth order Hamiltonian 
describes the motion of electrons 
along the $y$--direction, as defined 
in Fig. [\ref{RSVSVC}(a)], in the stripe potential. 
The unit cell is doubled along this 
direction due to the spin potential, 
as shown by the boxes in Fig. [\ref{RSVSVC}(a)]. 
It is convenient to introduce a pseudo--spinor 
operator, 
\begin{equation}
 \hat{\psi}_{xy}=\left(\begin{array}{c}
                        \psi_{xy}^B\\
                        \psi_{xy}^A
                       \end{array}
\right),
\end{equation}
where $y$ is now multiples of 2 lattice constants, and 
$\psi_{xy}^{A,B}$ is the electron annihilation 
operator. Then, for the column $x=n$, the zeroth 
order Hamiltonian is given by 
\begin{equation}
 \mathcal{H}_n^{(0)}=\sum_{p_y}
\hat{\psi}_{n,p_y}^{\dagger}
\hat{H}^{(0)}_{n,p_y}
\hat{\psi}_{n,p_y},
\end{equation}
with 
\begin{equation}
\hat{H}_{n,p_y}^{(0)}
= V^c_n+V^s_n\hat{\tau}_z-
2t^{\prime\prime}
\cos 2p_y-t[(1+\cos 2p_y)\hat{\tau}_x-\sin2p_y\hat{\tau}_y],
\label{original Hn}
\end{equation}
where $\hat{\tau}$'s are the Pauli matrices, and 
$V^c_0=-V^c_2=-2V_c$, $V^c_1=V^c_3=0$, $V^s_0=0$, 
$V^s_1=V^s_3=\sqrt{2}V$, and $V^s_2=-2V$. 
After a canonical transformation $\mathcal{T}$, 
which rotates the pseudo--spinor $\hat{\psi}$ in the 
pseudo--spin space first about the $\hat{\tau}_z$--axis 
by $p_y/2$ and then about the $\hat{\tau}_y$--axis 
by $\pi/2$, Eq. [\ref{original Hn}] becomes
\begin{equation}
\hat{H}_{n,p_y}^{(0)}= V^c_n+V^s_n\hat{\tau}_x-
2t^{\prime\prime}\cos 2p_y+2t\cos p_y\hat{\tau}_z.
\label{rotated Hn}
\end{equation}
The energy bands on the column $n=0$ are 
\begin{equation}
 E^0_{\pm}=-2V_c-2t^{\prime\prime}\cos 2p_y
\pm2t|\cos p_y|,
\end{equation}
with corresponding wave functions 
\begin{equation}
 |+>=\left(\begin{array}{c}
            \theta(\cos p_y)\\
            \theta(-\cos p_y)
           \end{array}
\right),\qquad
|->=\left(\begin{array}{c}
           \theta(-\cos p_y)\\
           \theta(\cos p_y)
          \end{array}
\right).
\end{equation}
For hole doping $x$ in the range $0.125<x<0.25$,
$\varepsilon^0(p_y)=E^0_-$. To leading order in $t/V$, 
the energy bands are 
$E^1_{\pm}=\pm \sqrt{2}V$ on the column $n=1$, 
$E_{\pm}^2=2V_c\pm 2V$ on the column $n=2$, and 
$E^3_{\pm}=\pm\sqrt{2}V$ on the column $n=3$. 
Fig. [\ref{RSVSVC}(b)] shows these energy bands 
for each column. 

The motion of electrons along the $x$--direction 
is described by 
\begin{equation}
 \mathcal{H}^X=\sum_{n,p_y}[\hat{\psi}_{n+1,p_y}^{\dagger}
H^+_{p_y}\hat{\psi}_{n,p_y}+h.c.]+\sum_{n,p_y}
[\hat{\psi}_{n+2,p_y}^{\dagger}H^{++}\hat{\psi}_{n,p_y}+h.c.],
\end{equation}
where, after the canonical transformation $\mathcal{T}$, 
\begin{equation}
 H^+_{p_y}=-t-2t^{\prime}\cos p_y\hat{\tau}_z,
\end{equation}
and 
\begin{equation}
 H^{++}=-t^{\prime\prime}.
\end{equation}

We calculate the matrix elements for electrons (holes) 
to hop from one charge stripe to a nearby stripe by 
perturbation theory using $\mathcal{H}^X$ as a perturbation. 
In the following, we first consider 
the case where there is no charge potential, 
and then consider the case where the 
charge potential $V_c$ is nonzero. 

\subsection{Spin Stripe Potential Only}

In the absence of the charge potential, the leading 
order terms in the matrix elements that describe 
electrons hopping among charge stripes are of order $1/V^2$. 
To this order, there are three possible processes, 
whose matrix elements are denoted 
by $M^A$, $M^B$ and $M^C$. $M^A$ is given by
\begin{equation}
 M^A=<-|H^{++}\frac{1}{E^0_--\hat{H}^{(0)}_{2,p_y}}H^{++}|->,
\end{equation}
and represents the hopping between stripes $n=0$ and $n=4$ by 
two $H^{++}$. 
$M^B$, which represents the hopping between stripes $n=0$ and 
$n=4$ by two $H^+_{p_y}$ and one $H^{++}$, is given by  
the sum of the three terms, 
\begin{equation}
 M^B_{211}=<-|H^{++}\frac{1}{E^0_--\hat{H}_{2,p_y}^{(0)}}
H^+_{p_y}\frac{1}{E^{(0)}_--\hat{H}_{3,p_y}^{(0)}}H^+_{p_y}|->,
\end{equation}
\begin{equation}
 M^B_{121}=<-|H^+_{p_y}\frac{1}{E^0_--\hat{H}_{1,p_y}^{(0)}}
H^{++}\frac{1}{E^0_--\hat{H}_{3,p_y}^{(0)}}H^+_{p_y}|->,
\end{equation}
and 
\begin{equation}
 M^B_{112}=<-|H^+_{p_y}\frac{1}{E^0_--\hat{H}_{1,p_y}^{(0)}}H^+_{p_y}
\frac{1}{E^0_--\hat{H}_{2,p_y}^{(0)}}H^{++}|->.
\end{equation}
$M^C$ represents the hopping between stripes $n=0$ and $n=8$ 
by four $H^{++}$, and is given by 
\begin{equation}
 M^C=<-|H^{++}\frac{1}{E^0_--\hat{H}_{2,p_y}^{(0)}}H^{++}|+>
\frac{1}{E^0_--E^0_+}<+|H^{++}\frac{1}{E^0_--\hat{H}_{6,p_y}^{(0)}}
H^{++}|->,
\end{equation}
where $V^s_6=-V^s_2$.  

To order $1/V^2$, the Hamiltonian $H$ 
in Eq. [\ref{general H}] is  
\begin{equation}
 H=\varepsilon^0(p_y)+f_1(p_y)\cos 4p_x
+f_2(p_y)\cos 8p_x,
\label{E large V}
\end{equation}
where 
\begin{equation}
\begin{split}
 f_1&=2(M^A+M^B_{112}+M^B_{121}+M^B_{211})\\
&=\frac{2tt^{\prime\prime}}{V^2}(2t^{\prime}+t^{\prime\prime})|\cos p_y|
-\frac{t^{\prime\prime}}{V^2}\bigl[(1-\sqrt{2})t^2+4(1+\sqrt{2})t^{\prime2}
\cos^2p_y\bigr],
\label{f1 large V}
\end{split}
\end{equation}
\begin{equation}
 f_2=2M^C=-\frac{t^{\prime\prime4}}{8tV^2|\cos p_y|},
\label{f2 large V}
\end{equation}
and we have neglected the hopping processes 
in which an electron hops from a charge stripe into 
the region between stripes with a large potential 
and hops back to the same stripe. 
These processes give corrections to $\varepsilon^0(p_y)$ 
of order $1/V^2$ with no $p_x$--dependence, and thus 
have no effects on $R_H$ to leading order. 
Eq. [\ref{E large V}] can also be obtained from 
third order perturbation calculation 
of Eq. [\ref{hamiltonian}], treating $\varepsilon_{p}$ 
as a perturbation.

\begin{figure}[htbp]
 \centering
\includegraphics[width=5in]{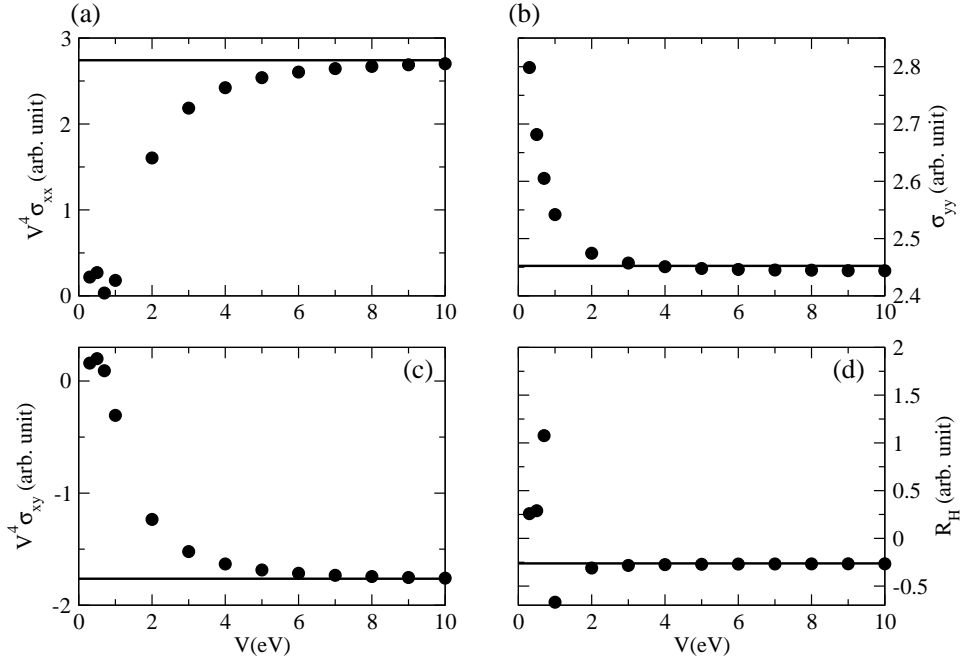}
\caption{\small{
Numerical results of the 
transport coefficients at large 
spin stripe potential $V$, 
for doping $x=0.125$, $V_c=0$, and for 
the canonical values of $t$, $t^{\prime}$, 
and $t^{\prime\prime}$. 
(a): $V^4\sigma_{xx}$ (dots) and 
$\mathcal{S}_{xx}$ (solid line). 
(b): $\sigma_{yy}$ (dots) and 
$\sigma_{yy}^{\infty}$ (solid line). 
(c): $V^4\sigma_{xy}$ (dots) and 
$\mathcal{S}_{xy}$ (solid line). 
(d): $R_H$ (dots) and $R_H^{\infty}$ (solid line). 
}}
\label{largeVasym}
\end{figure}

Substituting Eqs. [\ref{f1 large V},\ref{f2 large V}] 
into Eq. [\ref{sigxy large stripe}], 
we obtain the Hall conductivity $\sigma_{xy}$ 
to leading order in $1/V$ in the large--$V$ limit, 
\begin{equation}
 \sigma_{xy}=-\frac{16\pi\mathcal{A}}{V^4}
\Bigl\{f_1^2\bigl(\frac{m_0}{v_0}-\frac{1}{f_1}\frac{df_1}{dp_y}
\bigr)+4f_2^2\bigl(\frac{m_0}{v_0}-\frac{1}{f_2}
\frac{df_2}{dp_y}\bigr)\Bigr\}
\equiv \frac{1}{V^4}\mathcal{S}_{xy},
\label{sigxy large V}
\end{equation}
where all the quantities are evaluated at $p_0$ which is 
determined by the carrier density. Since both $\sigma_{xx}$ 
and $\sigma_{yy}$ are positive--definite, the sign 
of $R_H$ is determined by that of $\sigma_{xy}$. 
Similarly, substituting $f_1$ and $f_2$ into 
Eqs. [\ref{sigxx large stripe},\ref{sigyy large stripe}], 
we obtain the leading order terms of $\sigma_{xx}$ and 
$\sigma_{yy}$, 
\begin{equation}
 \sigma_{xx}=\frac{1}{V^4}\mathcal{B} \frac{16\pi}{|v_0|}
(f_1(p_0)^2+4f_2(p_0)^2)\equiv 
\frac{1}{V^4}\mathcal{S}_{xx},
\label{sigxx large V}
\end{equation}
and 
\begin{equation}
 \sigma_{yy}=\mathcal{B}2\pi|v_0|\equiv 
\sigma_{yy}^{\infty}. 
\label{sigyy large V}
\end{equation}
So $R_H$ approaches a constant as  
the stripe potential $V\to \infty$, 
\begin{equation}
 \lim_{V\to \infty}R_H=
\frac{\mathcal{S}_{xy}}{\mathcal{S}_{xx}
\sigma_{yy}^{\infty}}\equiv R_H^{\infty}. 
\label{RH large V}
\end{equation}
We observe that $\mathcal{S}_{xy}$, $\mathcal{S}_{xx}$, 
$\sigma_{yy}^{\infty}$ and $R_H^{\infty}$
are determined by the carrier density and the band 
parameters $t$, $t^{\prime}$, 
and $t^{\prime\prime}$. 
We perform numerical calculations of the 
conductivities $\sigma_{xx,xy,yy}$ 
and the Hall coefficient $R_H$ 
for the spin stripe potential $V$ up to 
10eV, doping $x=0.125$, $V_c=0$, and  
the canonical values of the band parameters: 
$t=0.38$eV, $t^{\prime}=0.32t$, and 
$t^{\prime\prime}=0.5t^{\prime}$. 
The results are shown as dots 
in Fig. [\ref{largeVasym}], 
where we compare these numerical results to 
the corresponding $V\to \infty$ limits 
(solid lines). We observe that the numerical 
results indeed approach the expected 
values. There are 
small discrepancies, which we attribute to  
the errors in calculating the chemical potential 
and in numerically finding 
the Fermi surface. For the parameters used 
here, $R_H^{\infty}<0$. In Sec. \ref{comparison}, 
we showed that for $V=0.3$eV, there is only open Fermi 
surface and $R_H>0$. Thus there is 
a change of sign in $R_H$ for $V>0.3$eV 
(roughly at $V=1$eV, Fig. [\ref{largeVasym}(d)]). 
This sign change can be understood qualitatively from 
Eqs. [\ref{general H},\ref{sigxy large stripe}], where 
we argued that $\sigma_{xy}$ changes sign when 
one of the $f_n$ goes through zero.  
We compared the Fermi surfaces for $V$ close to 
1eV, and found strong evidence that 
at least one of the $f_n$ in Eq. [\ref{general H}] 
changes sign.

\begin{figure}[htbp]
 \centering
\includegraphics[width=3in]{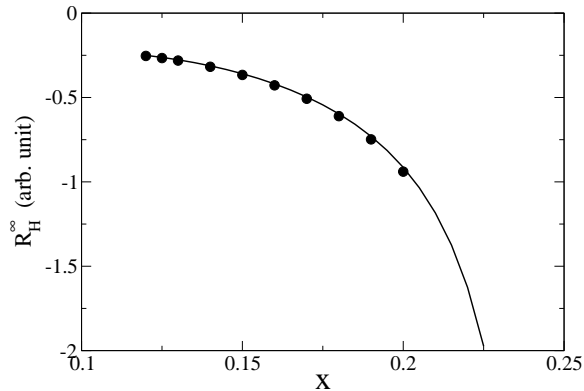}
\caption{\small{$R_H^{\infty}$ as a function of 
doping $x$ for $V_c=0$ and the canonical values of 
the band parameters.
Solid line: $R^{\infty}_H$ from Eq. [\ref{RH large V}]. 
Dots: numerical results for $V=10$eV.}}
\label{large V doping}
\end{figure}

\begin{figure}[htbp]
 \centering
\includegraphics[width=3in]{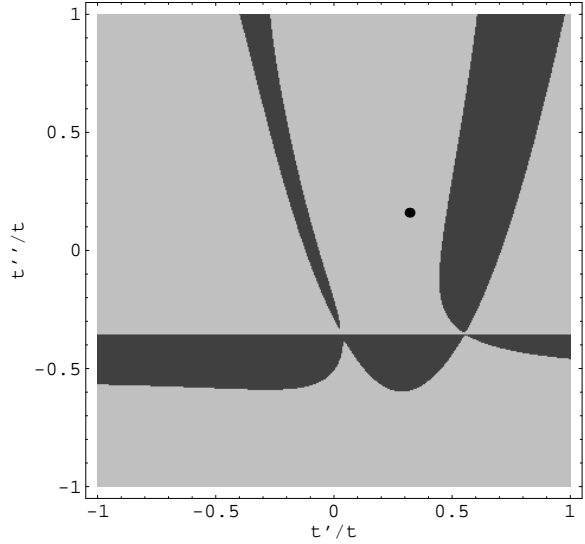}
\caption{\small{The sign map of $R_H^{\infty}$, as determined 
from Eq. [\ref{RH large V}], for $V_c=0$ and doping $x=0.125$. 
In the dark region $R_{H}^{\infty}>0$, 
and in the grey region $R_H^{\infty}<0$. The dot 
represents the point at 
$t^{\prime}/t=0.32$ and $t^{\prime\prime}/t=0.16$.}}
\label{signmap large V}
\end{figure}

We now study the dependence 
of $R_H^{\infty}$ on doping and band parameters. 
Fig. [\ref{large V doping}] shows 
$R_H^{\infty}$ as a function of doping, 
using the canonical values of the band 
parameters: $t^{\prime}/t=0.32$ and 
$t^{\prime\prime}/t=0.16$. 
We also calculated $R_H$  
as a function of doping numerically for 
$V=10$eV, shown as dots 
in Fig. [\ref{large V doping}]. 
There is a good agreement between 
the numerical results and $R_H^{\infty}$. 
We observe that for the canonical values of 
the band parameters and in the doping 
range $0.125<x<0.25$, $R_H^{\infty}<0$. 
The sign of $R_H^{\infty}$ as a function 
of the band parameters $t^{\prime}/t$ and 
$t^{\prime\prime}/t$ for doping 
$x=0.125$ is plotted in Fig. [\ref{signmap large V}], 
which shows that the area of the grey region 
where $R_H^{\infty}<0$ is much larger than 
that of the black region where $R_H^{\infty}>0$. 

\subsection{Coexistence of Charge Stripe Potential and Spin 
Stripe Potential}

In this subsection, we consider the case where 
$V_c=\alpha V$ with $\alpha<1/2$ and $V_c\gg t$. 
When $V_c$ is of the same order as $V$, 
the leading order term in the 
matrix element $M^A$ is of order $1/V$, and it is the only 
term at this order. To order $1/V$, the Hamiltonian in 
Eq. [\ref{general H}] takes the form 
\begin{equation}
 H=-2V_c-2t^{\prime\prime}\cos 2p_y-2t|\cos p_y|
+\tilde{f}_1\cos 4p_x,
\label{E large VSVC}
\end{equation}
where 
\begin{equation}
 \tilde{f}_1=2M^A=\frac{2\alpha}{1-4\alpha^2}
\frac{t^{\prime\prime2}}{V},
\label{f1 large VSVC}
\end{equation}
and terms of order $1/V$ and independent of $p_x$ are neglected.

\begin{figure}[htbp]
 \centering
\includegraphics[width=5in]{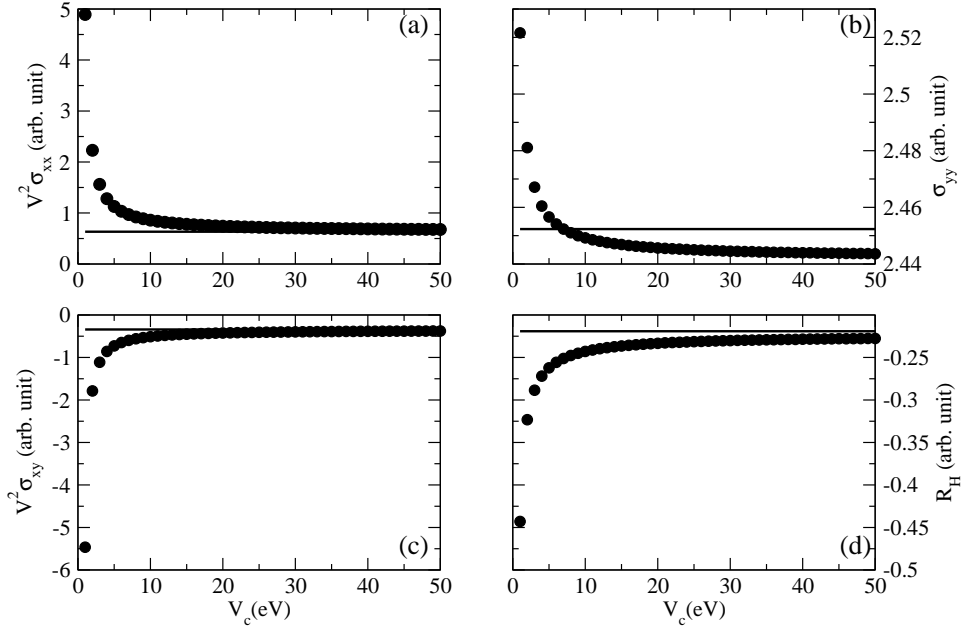}
\caption{\small{
Transport coefficients in the large spin and 
charge stripe potentials for $V=3V_c$, 
doping $x=0.125$, and the canonical values 
of the band parameters. 
(a): $V^2\sigma_{xx}$ (dots) and 
$\tilde{\mathcal{S}}_{xx}$ (solid line). 
(b): $\sigma_{yy}$ (dots) and 
$\sigma_{yy}^{\infty}$ (solid line). 
(c): $V^2\sigma_{xy}$ (dots) 
and $\tilde{\mathcal{S}}_{xy}$ (solid line). 
(d): $R_H$ (dots) and $\tilde{R}_H^{\infty}$ (solid line).}}
\label{large VSVC asym}
\end{figure}

Substituting Eq. [\ref{f1 large VSVC}] 
into Eq. [\ref{sigxy large stripe}], 
the Hall conductivity $\sigma_{xy}$ to leading order in $1/V$ 
is given by 
\begin{equation}
 \sigma_{xy}=-\frac{16\pi\mathcal{A}}{V^2}\frac{m_0}{v_0}
\tilde{f}_1^2\equiv \frac{1}{V^2}
\tilde{\mathcal{S}}_{xy},
\label{sigxy large VSVC}
\end{equation}
where we have used $d\tilde{f}_1/dp_y=0$. The longitudinal 
conductivies $\sigma_{xx}$ and $\sigma_{yy}$ are calculated in 
a similar manner, to leading order in $1/V$, 
\begin{equation}
 \sigma_{xx}=\frac{1}{V^2}\mathcal{B}\frac{16\pi}{|v_0|}
\tilde{f}_1^2\equiv \frac{1}{V^2}\tilde{\mathcal{S}}_{xx},
\label{sigxx large VSVC}
\end{equation}
and $\sigma_{yy}$ is given by Eq. [\ref{sigyy large V}]. 
So, $R_H$ approaches 
a constant in the $V\to \infty$ limit, 
\begin{equation}
\lim_{V\to \infty}R_H\equiv 
\tilde{R}_H^{\infty}=\frac{\tilde{\mathcal{S}}_{xy}}
{\tilde{\mathcal{S}}_{xx}\sigma_{yy}^{\infty}}\sim 
-\frac{m_0}{v_0}.
\label{RH large VSVC} 
\end{equation}
We see that $\tilde{\mathcal{S}}_{xy}$, 
$\tilde{\mathcal{S}}_{xx}$, 
$\sigma_{yy}^{\infty}$, and $\tilde{R}_H^{\infty}$  
are determined by the ratio $\alpha=V_c/V$, the hole doping 
$x$, and the band parameters $t$, $t^{\prime}$ and 
$t^{\prime\prime}$. 
The numerical results of the conductivities $\sigma_{xx,xy,yy}$ 
and the Hall coefficient $R_H$ for the charge stripe potential 
$V_c$ up to 50eV, the spin stripe potential $V=3V_c$, 
doping $x=0.125$, and the canonical values of the band 
parameters are shown in Fig. [\ref{large VSVC asym}], 
where we compare the results 
to the corresponding $V\to \infty$ limits.  
We observe that the numerical results approach 
the expected values, with small discrepancies 
which we attribute to the errors 
in calculating the chemical potential and in finding 
the Fermi surface numerically. For the parameters used 
here, $\tilde{R}_H^{\infty}<0$. 

\begin{figure}[htbp]
 \centering
\includegraphics[width=4in]{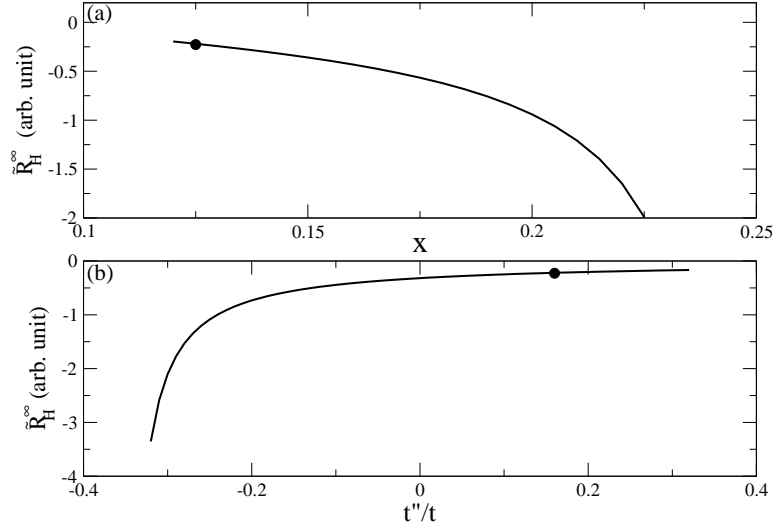}
\caption{\small{(a): $\tilde{R}_H^{\infty}$ 
as a function of doping $x$, 
for $V_c/V=1/3$ and $t^{\prime\prime}/t=0.16$. 
(b): $\tilde{R}_H^{\infty}$ as a function 
of $t^{\prime\prime}/t$, for $V_c=V/3$ and doping $x=0.125$. 
In both plots solid lines are results from 
Eq. [\ref{RH large VSVC}], and 
dots are from numerical calculation for $V_c=50$eV,  
$V=150$eV, doping $x=0.125$, $t^{\prime}/t=0.32$, and 
$t^{\prime\prime}/t=0.16$.}}
\label{RH VSVC x and t3}
\end{figure}

Eq. [\ref{RH large VSVC}] shows that 
$\tilde{R}_H^{\infty}$ is entirely determined by the 
curvature of the 1D band at $p_0$: $m_0/v_0$. 
We now study $\tilde{R}_H^{\infty}$ as a function of doping 
and the band parameters. 
In Fig. [\ref{RH VSVC x and t3}(a)], we plot 
$\tilde{R}_H$ as a function of 
doping $x$ for the canonical values of the 
band parameters, $t^{\prime}/t=0.32$ and 
$t^{\prime\prime}/t=0.16$. We see that 
for this set of band parameters, $\tilde{R}_H^{\infty}<0$ 
for doping $0.125<x<0.25$, because both $m_0$ and 
$v_0$ are positive in this doping range. 
In fact, it is easy to see that $\tilde{R}_H^{\infty}<0$ 
in this doping range as long as $t^{\prime\prime}/t>0$. 
Since $\varepsilon^0(p_y)$ is independent of $t^{\prime}$, 
we only need to study the effects of $t^{\prime\prime}/t$ on 
$\tilde{R}_H^{\infty}$. 
Fig. [\ref{RH VSVC x and t3}(b)] shows  
$\tilde{R}_H^{\infty}$ as a function of $t^{\prime\prime}/t$ 
for the doping $x=0.125$. We see that $\tilde{R}_H^{\infty}<0$ 
for $-0.32<t^{\prime\prime}/t<0.32$.

We mention that when the charge stripe potential $V_c\sim t$ 
and $V_c\ll V$, $\tilde{f}_1$ in 
Eq. [\ref{f1 large VSVC}] is of the same order as $f_1$ in 
Eq. [\ref{f1 large V}] and $f_2$ in Eq. [\ref{f2 large V}]. 
In this case, to leading order $1/V^2$, 
the Hamiltonian $H$ in Eq. [\ref{general H}] has the form 
of Eq. [\ref{E large V}]; the only difference is that the 
coefficient of $\cos 4p_x$ is now given by $f_1+\tilde{f}_1$. 
The Hall effect in this case is then similar to 
that in the $V_c=0$ case.

\section{discussion}
\label{discussion}

To conclude, we have considered the Hall effect in a stripe--ordered 
system. We found that the Hall effect $R_H$ shows complicated behavior 
as the spin stripe potential $V$ and/or the charge 
stripe potential $V_c$ are varied. For moderate values of $V$ and 
$V_c$, the behavior of $R_H$ can be understood 
as a result of the change of the 
Fermi surface topology, which is quite sensitive to the tuning of 
the stripe order potentials. In the strong coupling limit, the 
sign of $R_H$ was also found to depend on details. 

In a model with only static spin stripe potential $V$, $R_H$ first 
increases from a positive band value $R_H^0$, then decreases 
to negative values, and goes back to positive values, as 
increasing $V$ up to $\sim 1$eV, and then has a further sign change 
at unphysically large $V$. 
This initial increase and the subsequent change of 
sign qualitatively agrees with the experimental data. 
This is further supported from the model calculation in which 
$V$ is assumed to increase when decreasing doping from $x=0.24$, as 
shown in Fig. [\ref{rhplot}]. 
We mention that analogous calculations (not shown here) 
based on spiral order, do not produce a sign change. 
In a model with only static charge  stripe potential $V_c$, 
our calculation shows that 
$R_H$ increases from the band value until a sign change at 
$V_c\approx0.8$eV, after which the sign assumes the 
strong coupling limit electron--like value. 
This is qualitatively inconsistent with 
experimental data. 

When both the static spin stripe potential and the 
charge stripe potential are present, $R_H$ can be 
strongly enhanced or can be made negative 
by tuning $V$ and $V_c$, as shown in Fig. [\ref{vsvc}]. 
While both the spin stripe model and the $V$\&$V_c$ model 
produce a sign change in $R_H$, the mechanisms are different. 
In the spin stripe model, 
the sign change of $R_H$ is due to the electron pocket 
centered at $(0,\pi)$ and the elimination of the 
hole pockets centered at $(\pm\pi/8,\pi/2)$. In the 
spin and charge stripe model, the sign change is due to the 
merging of the hole pockets into open Fermi surface 
and the appearance of the electron pocket centered 
at $(0,0)$. 
Measurements directly probing 
the Fermi surfaces are required to distinguish these 
two scenarios. In our calculation, 
we found that the open Fermi surface 
can give either a positive ({\it i.e.} hole--like) contribution  
or a negative ({\it i.e.} electron--like) 
contribution to $\sigma_{xy}$. Under certain situations, 
especially when there is only open Fermi surface, 
this contribution, albeit small, is important, 
since the small $\sigma_{xx}$ would compensate the smallness of 
$\sigma_{xy}$ to give a large $|R_H|$; the $V=0.3$eV, $V_c=0$ point 
in Fig. [\ref{changeV}(d)] and the $V=0.25$eV, $V_c=0.7$eV point 
in Fig. [\ref{vsvc}(b)] are two examples. However, once there 
are electron or hole pockets, the contribution 
to $\sigma_{xy}$ from the open Fermi surface is negligible, 
and thus the sign of $R_H$ is fixed. 

We also considered the large stripe potential limit, in which 
the system is quasi--one dimensional, and the Fermi surface 
is open. We showed that analytical results of $R_H$ can 
be obtained in the limit $V\gg t$, both for $V\gg V_c$ and 
for $V>2V_c\gg t$. In this limit, $R_H$ depends on the carrier 
density, the electron band parameters, 
and the charge potential $V_c$, and its sign can be 
positive or negative. 

There remain discrepancies between experiment and theory. 
Experiment shows that $R_H$ at $x=0.2$ is about 
4 times larger than that at $x=0.24$, while our calculation 
only shows a factor of 2. However, the magnitude of $R_H$ 
depends crucially on the details of the Fermi surface. 
Angle dependence of the scattering rate\cite{Hussey08} 
(not considered here) may also be important. 
A systematic study of the doping dependence of 
the low temperature Hall effect, 
as was done on PCCO,\cite{Dagan04} would be 
helpful. But, the crucial generic result of our calculation 
is that the sign change of $R_H$ observed 
in Refs. [\onlinecite{Daou08},\onlinecite{Nakamura92}] 
appears to be strong 
evidence in favor of a spin--stripe order.

\begin{acknowledgements}
We thank the authors of Ref. [\onlinecite{Daou08}] for sharing 
their data in advance of publication, and M. R. Norman 
for helpful discussions. This work was supported by 
NSF--DMR--0705847.
\end{acknowledgements}



\end{document}